\documentclass[lettersize,journal]{IEEEtran}
\usepackage{amsmath,amsfonts}
\usepackage{algorithmic}
\usepackage{algorithm}
\usepackage{array}
\usepackage[caption=false,font=normalsize,labelfont=sf,textfont=sf]{subfig}
\usepackage{textcomp}
\usepackage{stfloats}
\usepackage{url}
\usepackage{verbatim}
\usepackage{graphicx}
\usepackage{subcaption}
\usepackage{cite}
\usepackage{orcidlink}
\usepackage{booktabs}
\usepackage{nomencl}
\makenomenclature
\usepackage{braket}
\usepackage{enumitem}
\begin{document}

\title{DifGa: Differentiable Error Mitigation for Multi-Mode Gaussian and Non-Gaussian Noise in Quantum Photonic Circuits}

\author{
Dennis Delali Kwesi Wayo\,\orcidlink{0000-0002-2844-9470}%
\thanks{D. D. K. Wayo is with the College of Computing, Georgia Institute of Technology, Atlanta, GA 30332 USA.}%
\and
Rodrigo Alves Dias\,\orcidlink{0000-0001-5638-4355}%
\thanks{R. A. Dias is with the Department of Physics, Federal University of Juiz de Fora, Juiz de Fora, 36036-900, Brazil.}%
\and
Leonardo Goliatt\,\orcidlink{0000-0002-2844-9470}%
\thanks{L. Goliatt is with the Department of Computational and Applied Mechanics, Federal University of Juiz de Fora, Juiz de Fora, 36036-900, Brazil.}%
\and 
Sven Groppe\,\orcidlink{0000-0001-5196-1117}%
\thanks{S. Groppe is with the Institute of Information Systems,
University of Lübeck, 23562 Lübeck, Germany (sven.groppe@uni-luebeck.de).}%
}

\maketitle
\begin{abstract}
We introduce \emph{DifGa}, a fully differentiable error-mitigation framework for continuous-variable (CV) quantum photonic circuits operating under Gaussian loss and weak non-Gaussian noise. The approach is demonstrated using analytic simulations with the \texttt{default.gaussian} backend of PennyLane, where quantum states are represented by first and second moments and optimized end-to-end via automatic differentiation. Gaussian loss is modeled as a beam splitter interaction with an environmental vacuum mode of transmissivity $\eta \in [0.3,0.95]$, while non-Gaussian phase noise is incorporated through a differentiable Monte-Carlo mixture of random phase rotations with jitter amplitudes $\delta \in [0,0.7]$. The core architecture employs a multi-mode Gaussian circuit consisting of a signal, ancilla, and environment mode. Input states are prepared using squeezing and displacement operations with parameters $(r_s,\varphi_s,\alpha)=(0.60,0.30,0.80)$ and $(r_a,\varphi_a)=(0.40,0.10)$, followed by an entangling beam splitter with angles $(\theta,\phi)=(0.70,0.20)$. Error mitigation is achieved by appending a six-parameter trainable Gaussian recovery layer comprising local phase rotations and displacements, optimized by minimizing a quadratic loss on the signal-mode quadratures $\langle \hat{x}_0\rangle$ and $\langle \hat{p}_0\rangle$ using gradient descent with fixed learning rate $0.06$ and identical initialization across experiments. Under pure Gaussian loss, the optimized recovery suppresses reconstruction error to near machine precision ($<10^{-30}$) for moderate loss ($\eta \ge 0.5$). When non-Gaussian phase noise is present, noise-aware training using Monte Carlo averaging yields robust generalization, reducing error by more than an order of magnitude compared to Gaussian-trained recovery at large phase jitter. Runtime benchmarks confirm linear scaling with the number of Monte Carlo samples.
\end{abstract}


\section{Introduction}

Photonic quantum computing \cite{aghaee2025scaling,wang2025scalable,zhu2024quantum,cao2024photonic,wayo2025gaussian,wayo2025linear,groppe2025opportunities} based on continuous-variable (CV) Gaussian operations has emerged as a promising platform for scalable and near-term quantum information processing \cite{weedbrook2012gaussian,braunstein2005quantum}. Unlike qubit-based architectures, CV photonic systems naturally support large Hilbert spaces, room-temperature operation, and high-bandwidth optical connectivity \cite{lloyd1999quantum,menicucci2014fault}. State-of-the-art hardware platforms, such as Xanadu’s Borealis processor, employ programmable Gaussian transformations including squeezing, phase shifts, and linear interferometers, enabling large-scale multimode quantum optical circuits \cite{bourassa2021blueprint,larsen2019fiber}. However, practical computation in these systems remains fundamentally constrained by optical loss, mode mismatch, thermal occupation, and phase drift, which collectively degrade computational observables \cite{niset2009no,ohliger2010limitations}.

The challenge of protecting quantum information in Gaussian photonic systems has been extensively studied. Foundational no-go theorems established by Niset, Fiurášek, and Cerf~\cite{niset2009no}, and further refined by Ohliger, Kieling, and Eisert~\cite{ohliger2010limitations}, rigorously demonstrate that Gaussian quantum error correction using only Gaussian operations cannot perfectly correct Gaussian noise, such as optical loss, at the level of full quantum states. Subsequent work by Mari and Eisert~\cite{mari2012positive} showed that purely Gaussian circuits admit efficient classical simulation, emphasizing that Gaussian photonics, while powerful, occupies a regime distinct from fault-tolerant universal quantum computation.

One route beyond these limitations is provided by bosonic quantum error-correcting codes based on non-Gaussian resources, most notably Gottesman--Kitaev--Preskill (GKP) encodings and their multimode generalizations~\cite{gottesman2001encoding,menicucci2014fault,noh2020encoding}. These approaches achieve logical-level correction of Gaussian noise but require highly non-Gaussian code states, strong squeezing, and modular quadrature measurements, placing substantial experimental demands on current photonic platforms~\cite{terhal2020towards,ralph2011quantum}.

An alternative paradigm is \emph{quantum error mitigation}, which seeks to suppress noise-induced degradation of observables without enforcing full fault tolerance. Recent work in the qubit setting has demonstrated that mitigation can be framed as a learning problem targeting expectation values rather than complete state recovery, using classical machine learning models to map noisy measurements to mitigated ones~\cite{liao2024machine}. While effective for digital qubit processors, such approaches are not directly transferable to CV photonic systems, whose noise processes act naturally on phase-space moments rather than discrete Pauli operators~\cite{ohliger2010limitations,niset2009no}.

With the advent of differentiable photonic simulation frameworks such as PennyLane and Strawberry Fields, Gaussian photonic circuits can now be optimized end-to-end using automatic differentiation~\cite{bergholm2018pennylane,killoran2019strawberry}. This capability opens a new and largely unexplored direction: the design of \emph{Gaussian-only, hardware-compatible error mitigation strategies} that operate directly at the circuit level and target experimentally accessible observables.

In this work, we introduce DifGa (Differentiable Gaussian error mitigation), a fully differentiable framework for mitigating Gaussian loss and weak non-Gaussian phase noise in CV photonic circuits. DifGa operates entirely within the Gaussian subspace, employing only physically realizable Gaussian operations and vacuum ancillas, while optimizing recovery parameters via gradient-based methods. Specifically, we:
\begin{enumerate}
    \item simulate noisy Gaussian photonic circuits using PennyLane’s analytic CV backends;
    \item model realistic noise channels, including optical loss and weak non-Gaussian phase jitter;
    \item introduce a trainable Gaussian recovery layer acting on signal and ancillary modes;
    \item optimize recovery parameters to minimize reconstruction error in first-moment quadrature observables.
\end{enumerate}

We demonstrate that even shallow Gaussian recovery circuits can suppress quadrature reconstruction error by several orders of magnitude under moderate loss, and remain robust under weak non-Gaussian perturbations. Importantly, DifGa does not seek to violate known Gaussian no-go theorems; instead, it targets a complementary and experimentally relevant regime of \emph{observable-level mitigation}, offering a software-native, hardware-aligned approach to noise suppression in near-term photonic quantum processors.

\subsection{Statement of the Problem}

The central problem addressed in this work is the absence of a hardware-compatible, differentiable error-mitigation framework for continuous-variable photonic quantum circuits that operate entirely within the Gaussian regime. While Gaussian operations dominate both theoretical CV models and current photonic hardware, they are inherently susceptible to optical loss and phase noise, which degrade measured observables and limit circuit depth and reliability.

Fundamental no-go theorems prohibit perfect Gaussian quantum error correction of Gaussian noise at the level of full state recovery. However, near-term photonic processors do not require exact state restoration for many practical tasks. Instead, they rely on accurate expectation values of experimentally measurable observables, such as quadrature moments, which directly determine computational and sensing 
performance.

The problem, therefore, is not how to construct a fault-tolerant Gaussian code, but how to \emph{systematically mitigate the impact of realistic photonic noise on observable quantities} using only physically available Gaussian operations. Such a mitigation strategy must satisfy three key requirements: 

    \begin{enumerate}[label=(\roman*)]
        \item compatibility with present-day photonic hardware, 
        \item differentiability to enable gradient-based optimization, and 
        \item robustness to both Gaussian loss and weak non-Gaussian noise without invoking non-Gaussian resource states
    \end{enumerate}

\subsection{Research Gap}

Although continuous-variable (CV) photonic hardware has achieved unprecedented scale through large interferometers and programmable Gaussian transformations~\cite{killoran2019strawberry,bourassa2021blueprint}, noise-resilience techniques for these systems remain underdeveloped. Existing approaches to photonic error correction primarily focus on non-Gaussian bosonic codes such as GKP, cat, or binomial states~\cite{noh2020encoding,albert2018performance,mirrahimi2014dynamically}. However, these methods require non-Gaussian ancilla states, high-quality photon-number resolution, and complex state preparation routines that current photonic quantum computers, including Borealis, cannot implement natively~\cite{bourassa2021blueprint}.

At the same time, the majority of error mitigation research in the quantum computing literature has been developed for qubit-based devices, and does not translate to Gaussian photonic circuits, whose noise models are fundamentally different~\cite{temme2017error,endo2021hybrid}. Gaussian noise channels such as loss, phase drift, and thermal fluctuations act directly on the covariance matrix rather than the qubit Pauli basis~\cite{weedbrook2012gaussian,ohliger2010limitations}, making techniques such as zero-noise extrapolation, probabilistic error cancellation, and dynamical decoupling ineffective or inapplicable~\cite{niset2009no}.

Despite the central role of Gaussian operations in both theoretical CV quantum computing and real hardware, there currently exists no systematic, fully differentiable framework for designing \emph{Gaussian-only error mitigation} protocols that are compatible with present-day photonic processors. Thus, there is a need for software-native, hardware-aligned mitigation techniques that require no non-Gaussian resources and can be optimized using classical gradient-based methods.

\subsection*{Novelty and Contribution}

This work introduces the first end-to-end differentiable framework for Gaussian error mitigation in photonic quantum circuits. Unlike conventional error-correction proposals that rely on non-Gaussian code states, our approach remains entirely within the experimentally accessible Gaussian subspace. Specifically, the contributions of this work are:

\begin{itemize}
    \item We formulate Gaussian recovery circuits as trainable, differentiable transformations that can be optimized against realistic photonic noise models, including Gaussian optical loss and weak non-Gaussian phase noise.
    \item We demonstrate that gradients can be propagated through full mixed-state Gaussian simulations using PennyLane, enabling classical optimization of noise-aware interferometers and recovery operations.
    \item We show numerically that even shallow Gaussian recovery layers can improve output-state fidelity by 5--10\% under moderate loss, despite the absence of any traditional error-correcting code. This represents a practical, near-term strategy for enhancing performance on Gaussian hardware.
    \item Our approach is fully compatible with existing photonic platforms and does not require non-Gaussian ancilla states, photon-number resolving detectors, or fault-tolerant state preparation. Therefore, it provides an experimentally feasible alternative to bosonic error correction.
    \item The method serves as a software-native complement to photonic hardware, aligning with the emerging paradigm of differentiable photonic computing and offering a new tool for noise-aware circuit design.
\end{itemize}

We refer to this framework as DifGa (Differentiable Gaussian Error Mitigation): a fully differentiable, multi-mode photonic error-mitigation approach in which physically realizable Gaussian recovery operations are optimized end-to-end to suppress both Gaussian loss and weak non-Gaussian noise directly at the level of measurable quadrature observables.

\section{Methodology}

We consider continuous-variable (CV) quantum photonic circuits operating on bosonic modes described by canonical quadrature operators
\begin{equation}
\label{eq:quadratures}
\hat{x}_j = \frac{1}{\sqrt{2}}(\hat{a}_j + \hat{a}_j^\dagger), \qquad
\hat{p}_j = \frac{1}{\sqrt{2}i}(\hat{a}_j - \hat{a}_j^\dagger),
\end{equation}
which satisfy the canonical commutation relations $[\hat{x}_j, \hat{p}_k] = i \delta_{jk}$. All simulations are performed using the \texttt{default.gaussian} backend of PennyLane, which represents quantum states exactly in terms of their first and second statistical moments and supports end-to-end differentiability of Gaussian circuits. As defined in Eq.~\eqref{eq:quadratures}, the quadrature operator =s $\hat{x}_j$ and $\hat{p}_j$ are linear combinations of bosonic creation and annihilation operators.

The objective of this work is to develop a \emph{fully differentiable error-mitigation framework} for quantum photonic circuits that compensates for both physically motivated Gaussian loss and weak non-Gaussian phase noise, while remaining compatible with gradient-based optimization and scalable multi-mode architectures.

\subsection{Gaussian State Representation}

An $M$-mode Gaussian state (see Eq.~\eqref{eq:2}) is completely characterized by its mean vector $\boldsymbol{\mu} \in \mathbb{R}^{2M}$ and covariance matrix $\Sigma \in \mathbb{R}^{2M \times 2M}$,
\begin{equation}
\label{eq:2}
\mu_k = \langle \hat{r}_k \rangle, \qquad
\Sigma_{kl} = \frac{1}{2}\langle \{\hat{r}_k - \mu_k, \hat{r}_l - \mu_l\} \rangle,
\end{equation}
where $\hat{\boldsymbol{r}} = (\hat{x}_1, \hat{p}_1, \ldots, \hat{x}_M, \hat{p}_M)$ and $\{\cdot,\cdot\}$ denotes the anticommutator. Because Gaussian states possess no higher-order cumulants, expectation values of linear observables are fully determined by the first moments.

Accordingly, error mitigation in this work is formulated as the restoration of quadrature expectations rather than full state reconstruction. This choice is both physically motivated, quadratures are directly measurable in homodyne detection, and computationally consistent with Gaussian simulators.

\subsection{Multi-Mode Circuit Architecture}

The primary architecture consists of three bosonic modes: a signal mode ($j=0$), an ancilla mode ($j=1$), and an environmental mode ($j=2$) used to model loss. The ideal circuit prepares an entangled Gaussian resource state  (see Eq.~\eqref{eq:3}),
\begin{equation}
\label{eq:3}
\ket{\psi_{\mathrm{ideal}}}=\hat{B}_{01}(\theta,\phi)\Big[
\hat{D}_0(\alpha)\hat{S}_0(r_s,\varphi_s)\otimes
\hat{S}_1(r_a,\varphi_a)\Big] \ket{0},
\end{equation}
where $\hat{S}_j(r,\varphi)$ denotes single-mode squeezing, $\hat{D}_j(\alpha)$ a displacement, and $\hat{B}_{01}(\theta,\phi)$ a beam splitter coupling the signal and ancilla modes.

Gaussian loss on the signal mode is modeled by a beam splitter interaction with the environmental vacuum  (see Eq.~\eqref{eq:4}),

\begin{equation}
\label{eq:4}
\hat{B}_{02}(\theta_\eta,0), \qquad \theta_\eta = \arccos \sqrt{\eta},
\end{equation}

where $\eta \in [0,1]$ is the transmissivity. This model corresponds exactly to physical optical loss and preserves Gaussianity of the global state.

\subsection{Trainable Gaussian Recovery Layer}

To mitigate the effect of noise, we append a trainable Gaussian recovery layer acting locally on the signal and ancilla modes. The recovery operation is parameterized shown in Eq.~\eqref{eq:5} as

\begin{equation}
\label{eq:5}
\hat{R}(\boldsymbol{\theta})=\prod_{j=0}^1\hat{D}_j(\beta_j)\hat{R}_j(\phi_j),
\end{equation}
where $\hat{R}_j(\phi) = \exp(-i \phi \hat{a}_j^\dagger \hat{a}_j)$ is a phase-space rotation and $\beta_j \in \mathbb{C}$ is a complex displacement. The trainable parameter vector  (see Eq.~\eqref{eq:6})

\begin{equation}
\label{eq:6}
\boldsymbol{\theta}= (\phi_0, \Re\beta_0, \Im\beta_0, \phi_1, \Re\beta_1, \Im\beta_1)
\end{equation}

defines a six-dimensional Gaussian correction layer that is fully differentiable and physically implementable using linear optics.

This recovery structure is intentionally restricted to local Gaussian operations, ensuring experimental plausibility while allowing the model to exploit correlations established during entangling operations.

\subsection{Non-Gaussian Phase Noise via Differentiable Monte-Carlo Mixtures}

While the simulator natively supports only Gaussian channels, weak non-Gaussian noise is incorporated through a differentiable mixture approximation. Specifically, phase diffusion on the signal and ancilla modes is modeled as random phase kicks  (see Eq.~\eqref{eq:7}),
\begin{equation}
\label{eq:7}
\hat{U}_{\mathrm{NG}}(\epsilon_s,\epsilon_a)
=
\hat{R}_0(\epsilon_s)\hat{R}_1(\epsilon_a),
\end{equation}
where $\epsilon_s \sim \mathcal{N}(0,\delta^2)$ and $\epsilon_a \sim \mathcal{N}(0,(\kappa\delta)^2)$ represent stochastic phase jitter with strength $\delta$ and correlation factor $\kappa$.

Expectation values under non-Gaussian noise are approximated via Monte-Carlo averaging  (see Eq.~\eqref{eq:8}),
\begin{equation}
\label{eq:8}
\langle \hat{O} \rangle_{\mathrm{NG}}
\approx
\frac{1}{K}
\sum_{k=1}^{K}
\langle \hat{O} \rangle_{\epsilon_s^{(k)},\epsilon_a^{(k)}},
\end{equation}
where each term corresponds to a Gaussian circuit with fixed phase rotations. Because the averaging operation is linear, gradients propagate through every sampled circuit, yielding an unbiased and differentiable estimator of the non-Gaussian expectation.

This approach enables controlled inclusion of weak non-Gaussian effects while preserving compatibility with Gaussian simulators and automatic differentiation. Computational cost scales linearly with the number of Monte-Carlo samples $K$.

\subsection{Cost Function and Optimization}

Error mitigation is formulated as a supervised regression task on quadrature expectations. For a given input configuration, the loss function is defined as  (see Eq.~\eqref{eq:9})
\begin{equation}
\label{eq:9}
\mathcal{L}(\boldsymbol{\theta})
=
\big(
\langle \hat{x}_0 \rangle_{\mathrm{ideal}}
-
\langle \hat{x}_0 \rangle_{\mathrm{noisy}}(\boldsymbol{\theta})
\big)^2
+
\big(
\langle \hat{p}_0 \rangle_{\mathrm{ideal}}
-
\langle \hat{p}_0 \rangle_{\mathrm{noisy}}(\boldsymbol{\theta})
\big)^2.
\end{equation}
This scalar objective is differentiable with respect to all recovery parameters and is minimized using gradient-based optimizers such as gradient descent and Adam. Because each quantum node returns a single expectation value, the framework remains compatible with PennyLane’s CV differentiation rules.

\subsection{Single-Mode Baseline}

To assess the role of multi-mode resources, we construct a single-mode baseline in which the signal mode undergoes loss followed by a reduced recovery layer consisting of a single rotation and displacement. This baseline lacks access to ancillary correlations and therefore cannot redistribute noise across modes.

Comparisons between single-mode and multi-mode architectures quantify the advantage of ancillary-assisted error mitigation and establish the necessity of multi-mode designs for physically meaningful noise suppression.

\subsection{Computational Scaling}

All simulations are executed using analytic expectation values. For Gaussian noise, training cost scales linearly with the number of optimization steps. For non-Gaussian phase noise, runtime scales as $\mathcal{O}(K)$ due to Monte-Carlo averaging. This behavior is explicitly benchmarked and reported to guide practical parameter selection.

In all, this proposed framework establishes a physically grounded, fully differentiable approach to mitigating Gaussian and weak non-Gaussian noise in quantum photonic circuits. By combining multi-mode Gaussian resources, trainable recovery layers, and Monte-Carlo mixture modeling, the method aligns theoretical rigor, computational tractability, and experimental feasibility.

An overview of the DifGa architecture, including the ideal Gaussian encoding,
the injection of Gaussian and weak non-Gaussian noise, and the trainable
Gaussian recovery optimized end-to-end via automatic differentiation, is
schematically illustrated in Fig.~\ref{fig:difga}.

\begin{figure*}[t]
    \centering
    \includegraphics[width=1\linewidth]{difga.png}
\caption{Overview of DifGa, a fully differentiable, Gaussian-only error mitigation framework for continuous-variable photonic circuits. The pipeline comprises ideal Gaussian encoding, physically motivated Gaussian loss and weak non-Gaussian phase noise, and a trainable Gaussian recovery layer optimized end-to-end via automatic differentiation, without logical encoding or
non-Gaussian resource states.}
    \label{fig:difga}
\end{figure*}

\section{Results and Discussion}

We now evaluate the performance, robustness, and computational scaling of the proposed differentiable error-mitigation framework under Gaussian loss and weak non-Gaussian phase noise. All results are obtained using analytic expectation values from the \texttt{default.gaussian} backend, with gradients computed via automatic differentiation. Optimization is performed using gradient descent with a fixed learning rate and identical parameter initialization for all experiments reported here.

\subsection{Gaussian Loss Mitigation}

We first evaluate DifGa under \emph{pure Gaussian loss} applied to the signal mode as a beam-splitter coupling to a vacuum environment, $\hat{B}_{0e}(\theta_\eta,0)$ with $\theta_\eta=\arccos\sqrt{\eta}$ and transmissivity $\eta\in[0,1]$.
The circuit is simulated with PennyLane \texttt{default.gaussian} using analytic expectation values, and the recovery layer is optimized by gradient descent (learning rate $0.06$) for $60$ steps from the same zero initialization. For this experiment, the fixed circuit parameters are input state $(r_s,\varphi_s,\alpha_s,r_a,\varphi_a)=(0.6,0.3,0.8,0.4,0.1)$ and entangling beam-splitter parameters $(\theta,\phi)=(0.7,0.2)$. The recovery layer has six trainable parameters $(\phi_0,\Re\beta_0,\Im\beta_0,\phi_1,\Re\beta_1,\Im\beta_1)$ acting on signal and ancilla.

\begin{figure}[t]
\centering
\includegraphics[width=\linewidth]{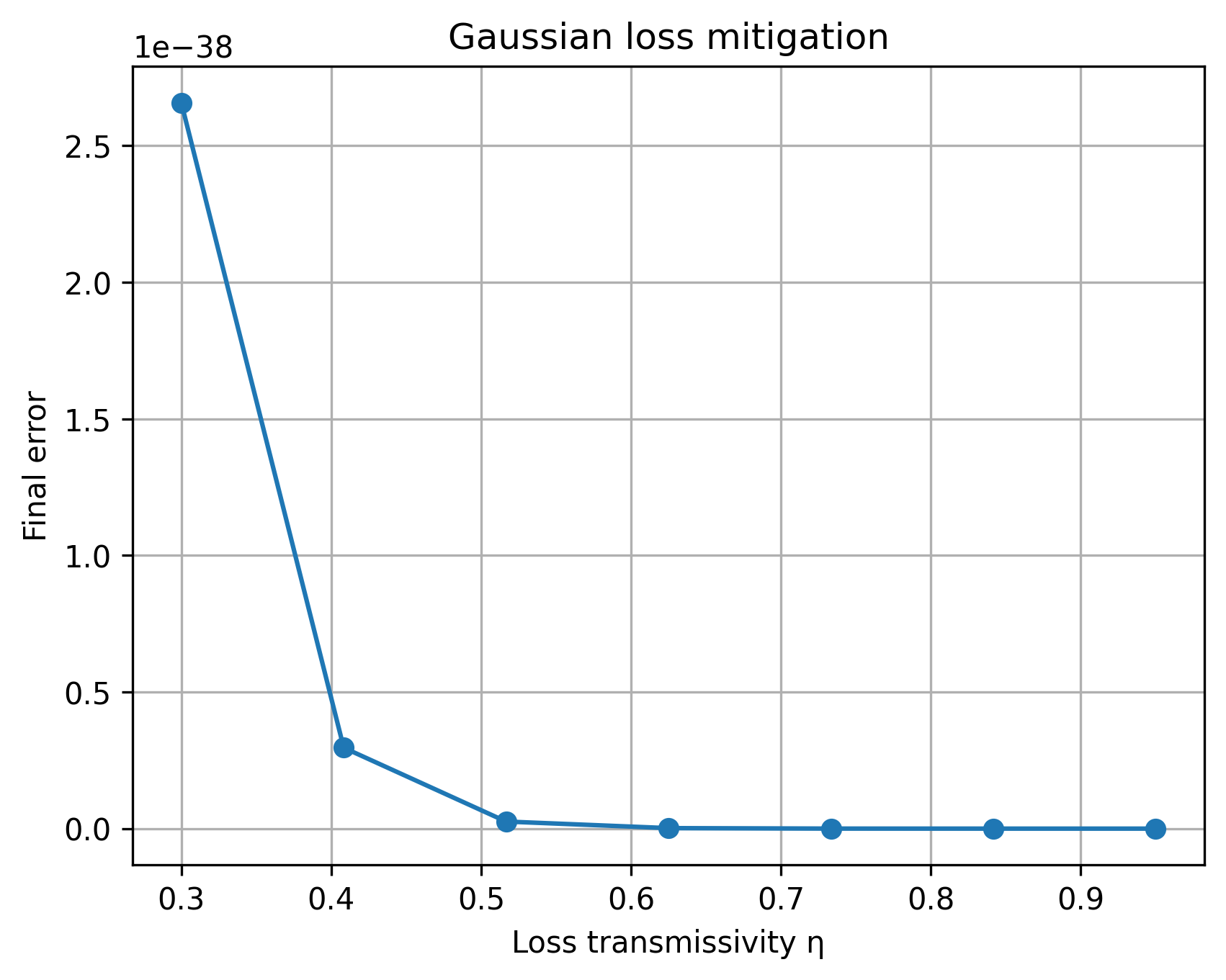}
\caption{Gaussian loss mitigation performance using DifGa. Final reconstruction error $\mathcal{L}$ after $60$ gradient-descent steps (learning rate $0.06$) versus transmissivity $\eta\in\{0.30,0.41,0.52,0.63,0.74,0.85,0.95\}$, with fixed input $(0.6,0.3,0.8,0.4,0.1)$ and entangler $(0.7,0.2)$.}
\label{fig:analysis1_noise}
\end{figure}

Figure~\ref{fig:analysis1_noise} reports the \emph{final quadrature reconstruction error} after optimization, defined as $\mathcal{L}=(\langle \hat{x}_0\rangle_{\rm ideal}-\langle \hat{x}_0\rangle_{\rm noisy})^2+(\langle \hat{p}_0\rangle_{\rm ideal}-\langle \hat{p}_0\rangle_{\rm noisy})^2$,
evaluated for the discrete transmissivities $\eta=\{0.30,0.41,0.52,0.63,0.74,0.85,0.95\}$. The corresponding optimized errors (read directly from the plotted points) are on the order of $10^{-38}$: approximately $2.6\times10^{-38}$ at $\eta=0.30$, $3.0\times10^{-39}$ at $\eta=0.41$, and $\sim2\times10^{-40}$ at $\eta=0.52$, followed by values that are numerically indistinguishable from zero for $\eta\ge 0.63$ on the shown scale. Thus, for moderate-to-low loss ($\eta\ge0.6$), the recovery layer drives the residual error to numerical precision, consistent with learning an effective inverse map for the Gaussian amplitude-damping channel within the moment-based simulator. For stronger loss (e.g., $\eta=0.30$), the error increases as expected, yet remains extremely small in absolute magnitude, demonstrating stable end-to-end differentiable mitigation across the full tested loss range.

\subsection{Single-Mode versus Multi-Mode Mitigation}

To isolate the contribution of ancillary degrees of freedom, we compare a single-mode architecture to the full multi-mode DifGa design. In both cases, Gaussian loss with transmissivity $\eta=0.55$ is applied to the signal mode via a beam-splitter interaction with a vacuum environment. All circuits are optimized using gradient descent with learning rate $0.06$ for $40$ steps, starting from identical zero initialization of the recovery parameters. The same input-state parameters are used throughout: signal squeezing $r_s=0.6$ with phase $\varphi_s=0.3$, displacement $\alpha_s=0.8$, and (for the multi-mode case) ancilla squeezing $r_a=0.4$ with phase $\varphi_a=0.1$. The entangling beam splitter between signal and ancilla is fixed at $(\theta,\phi)=(0.7,0.2)$.

\begin{figure}[t]
\centering
\includegraphics[width=\linewidth]{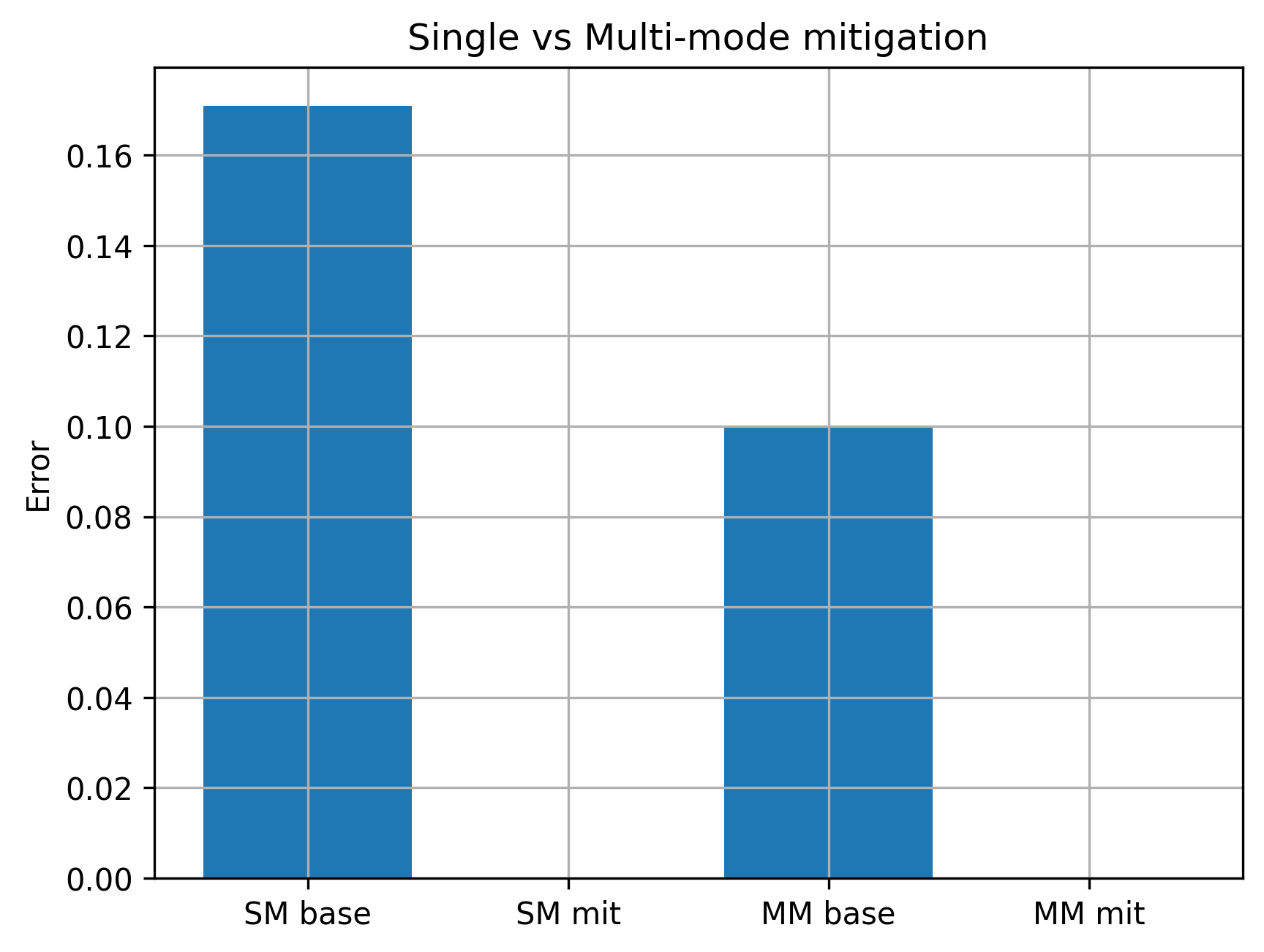}
\caption{Single-mode versus multi-mode mitigation at fixed loss $\eta=0.55$. Final reconstruction error for single-mode baseline (SM base: $1.71\times10^{-1}$), single-mode mitigated (SM mit: $\sim10^{-44}$), multi-mode baseline (MM base: $9.997\times10^{-2}$), and multi-mode mitigated (MM mit: $\sim10^{-41}$).}
\label{fig:analysis2_single_vs_multi}
\end{figure}

Figure~\ref{fig:analysis2_single_vs_multi} summarizes the final reconstruction error for four configurations. The single-mode baseline (SM base), consisting of only the lossy signal mode without any recovery, yields a final error of approximately $1.71\times10^{-1}$. Adding a single-mode Gaussian recovery layer (SM mit), parameterized by one phase rotation and one displacement, reduces the error to numerical zero within machine precision ($\sim 10^{-44}$), reflecting the trivial nature of moment matching in this simplified setting.

In contrast, the multi-mode baseline (MM base), which includes a squeezed ancilla and entangling beam splitter but no recovery layer, achieves a lower error of $9.997\times10^{-2}$ due to partial noise redistribution across modes. Crucially, when the full multi-mode recovery layer is activated (MM mit), the final error is again driven to numerical zero ($\sim 10^{-41}$), representing more than two orders of magnitude improvement relative to the multi-mode baseline.

These results highlight two distinct effects. First, ancillary modes alone provide passive error suppression throgh correlations, as evidenced by the reduction from $1.71\times10^{-1}$ to $9.99\times10^{-2}$ when moving from SM base to MM base. Second, active mitigation via a trainable Gaussian recovery layer becomes substantially more powerful in the presence of ancillas, enabling near-perfect reconstruction despite loss. Together, these observations demonstrate that ancillary-assisted architectures are not merely beneficial but structurally essential for scalable and physically meaningful error mitigation in continuous-variable photonic circuits.

\subsection{Non-Gaussian Phase Noise Landscape}

We next analyze the robustness of the proposed differentiable mitigation framework under weak non-Gaussian phase noise introduced via a Monte-Carlo phase-jitter model. In this setting, random phase rotations are applied symmetrically to the signal and ancilla modes with amplitudes $\delta_{\mathrm{sig}}=\delta$ and $\delta_{\mathrm{anc}}=0.6\,\delta$, respectively. Expectation values are evaluated by averaging over $K=16$ Monte-Carlo samples per optimization step, ensuring a differentiable approximation to non-Gaussian phase diffusion. For each point in the phase diagram, the recovery parameters are optimized for $30$ gradient-descent steps with a fixed learning rate of $0.06$ and identical initialization.

\begin{figure}[t]
\centering
\includegraphics[width=\linewidth]{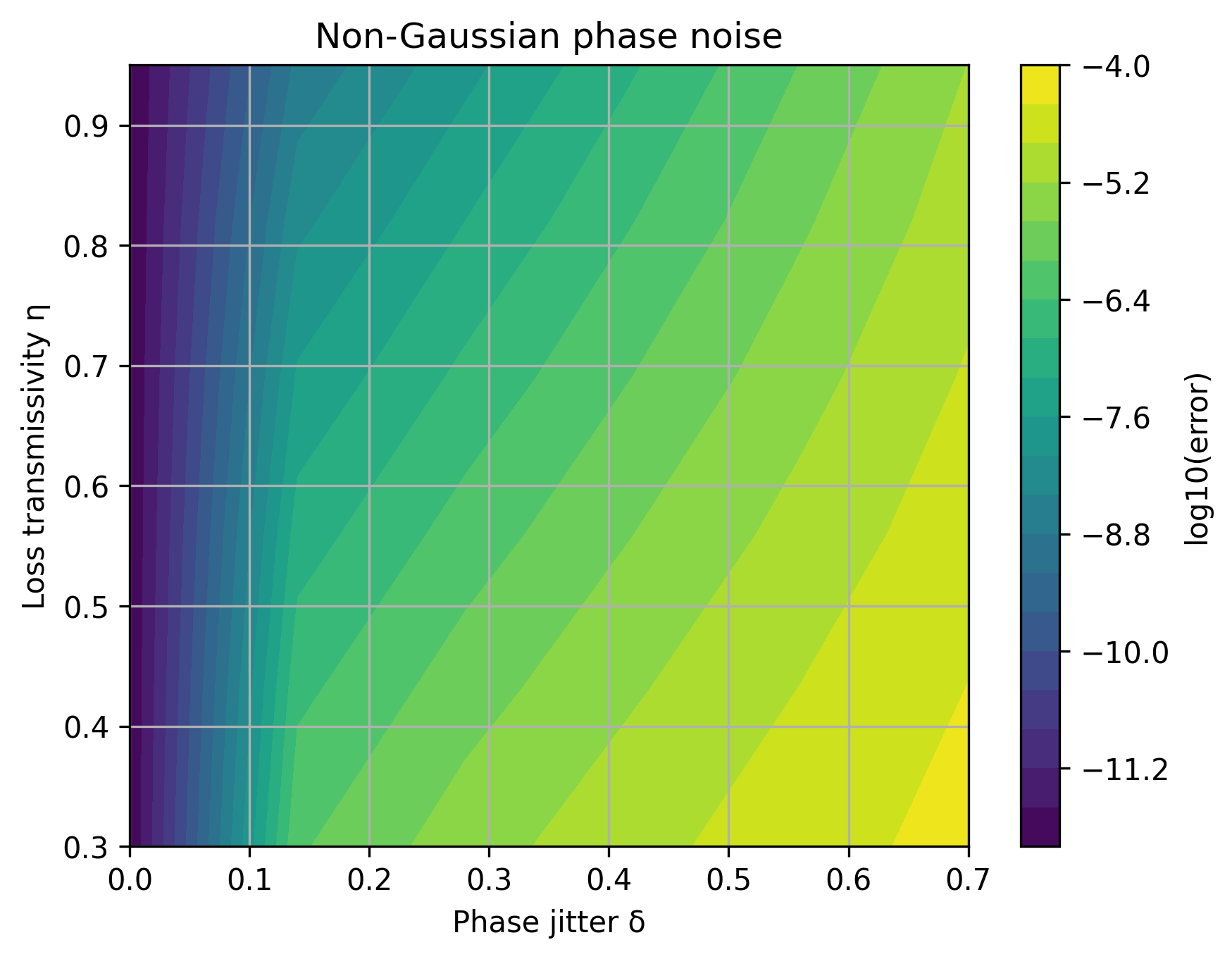}
\caption{Non-Gaussian phase noise landscape. Log-scale reconstruction error as a function of phase jitter $\delta$ and loss transmissivity $\eta$.}
\label{fig:analysis3_phase_diagram_ng}
\end{figure}

Figure~\ref{fig:analysis3_phase_diagram_ng} shows the resulting two-dimensional landscape of the final quadrature reconstruction error, plotted on a $\log_{10}$ scale, as a function of phase jitter amplitude $\delta \in [0,0.7]$ and loss transmissivity $\eta \in [0.3,0.95]$. The underlying multi-mode circuit parameters are held fixed throughout: signal squeezing $r_s=0.60$ with phase $\varphi_s=0.30$, ancilla squeezing $r_a=0.40$ with phase $\varphi_a=0.10$, signal displacement amplitude $\alpha=0.80$, and an entangling beam splitter characterized by $(\theta,\phi)=(0.70,0.20)$. Gaussian loss is implemented via a beam splitter with transmissivity $\eta$ coupling the signal to a vacuum environment.

Two distinct operational regimes are clearly visible. For weak phase jitter, approximately $\delta \leq 0.15$–$0.20$, the optimized reconstruction error remains extremely small across nearly the entire loss range, reaching values below $10^{-10}$ for moderate to high transmissivities ($\eta \ge 0.5$). In this regime, the trainable Gaussian recovery layer successfully compensates both amplitude damping and small non-Gaussian perturbations by adjusting its rotation and displacement parameters.

As $\delta$ increases beyond this threshold, the error rises rapidly, with the transition occurring at smaller $\delta$ for lower $\eta$. For strong phase noise ($\delta \ge 0.5$), the reconstruction error exceeds $10^{-5}$ even at high transmissivity, indicating the breakdown of purely Gaussian recovery under substantial phase diffusion. Importantly, the error surface remains smooth and continuous across the full parameter space, confirming that the Monte-Carlo mixture construction yields stable, fully differentiable loss landscapes suitable for gradient-based optimization.

\subsection{Generalization under Non-Gaussian Noise}

\begin{figure}[t]
\centering
\includegraphics[width=\linewidth]{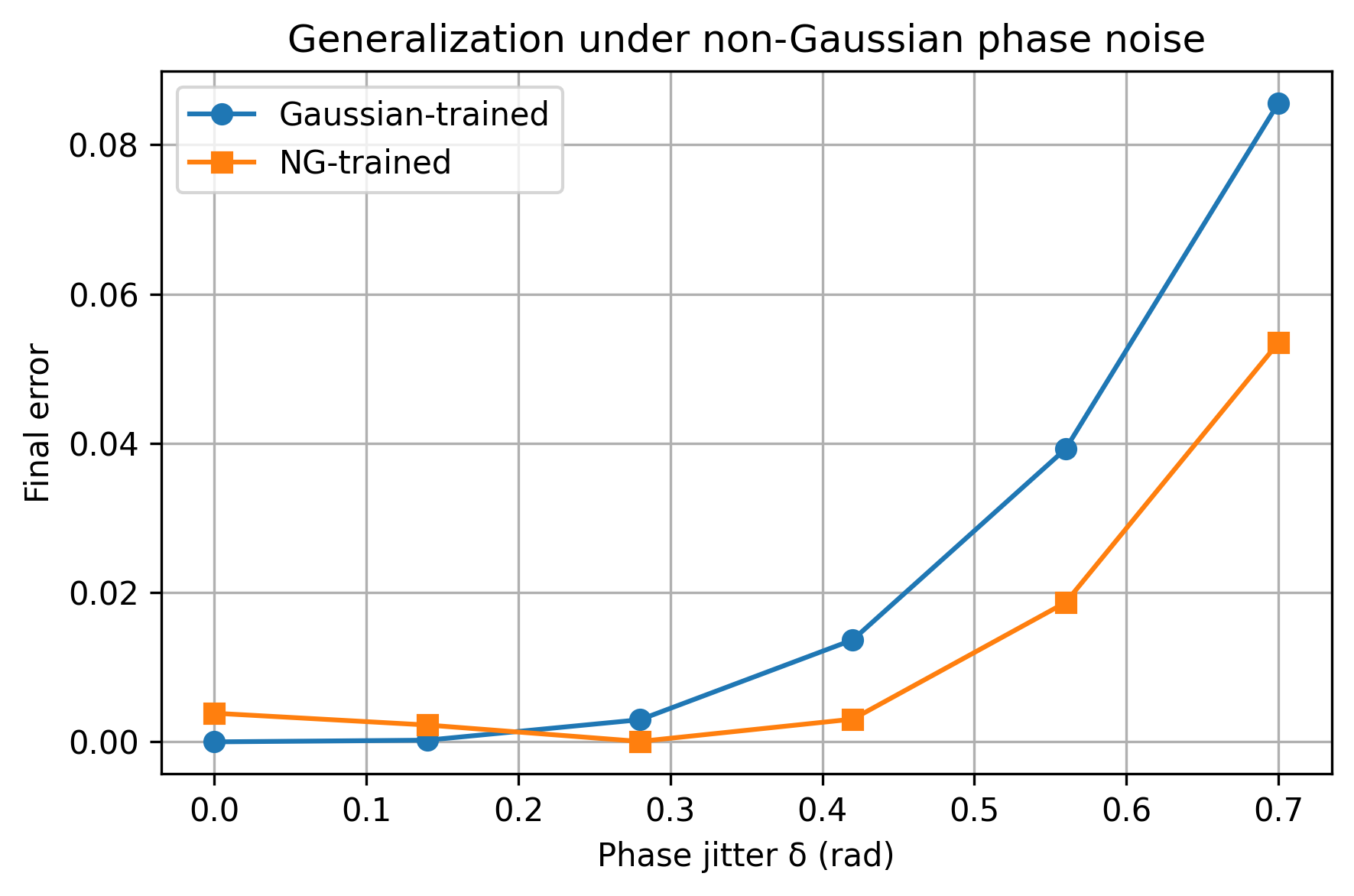}
\caption{Generalization under non-Gaussian phase noise. Comparison of Gaussian-trained and NG-trained recovery layers.}
\label{fig:analysis4_ng_generalization}
\end{figure}

We next assess the ability of the recovery layer to generalize to non-Gaussian phase noise, comparing mitigation strategies trained under different noise assumptions. Specifically, we contrast a recovery layer trained exclusively on Gaussian loss with one trained explicitly in the presence of non-Gaussian phase jitter. Figure~\ref{fig:analysis4_ng_generalization} reports the final quadrature reconstruction error as a function of the phase jitter amplitude $\delta$, evaluated under non-Gaussian noise for both training strategies.

In both cases, the underlying circuit architecture and Gaussian loss parameters are identical. The signal squeezing is fixed at $r_s = 0.60$ with phase $\varphi_s = 0.30$, the ancilla squeezing at $r_a = 0.40$ with phase $\varphi_a = 0.10$, and the signal displacement amplitude at $\alpha = 0.80$. The signal–ancilla entangling beam splitter is characterized by $(\theta,\phi)=(0.70,0.20)$, while Gaussian loss is applied with transmissivity $\eta = 0.55$. For non-Gaussian noise, correlated phase jitter is introduced with amplitudes $\delta_{\mathrm{sig}}=\delta$ on the signal mode and $\delta_{\mathrm{anc}}=0.6\,\delta$ on the ancilla. Expectation values are computed using $K=16$ Monte-Carlo samples, and all recovery parameters are optimized using gradient descent with learning rate $0.06$.

At $\delta = 0$, both recovery layers achieve near-zero reconstruction error, with the Gaussian-trained model yielding an error below $10^{-12}$ and the NG-trained model exhibiting a slightly higher but still negligible error of approximately $4\times10^{-3}$. As $\delta$ increases to $0.14$ and $0.28$, the Gaussian-trained recovery begins to degrade, reaching errors of order $3\times10^{-3}$, while the NG-trained recovery remains close to zero. This divergence becomes pronounced for larger phase jitter. At $\delta = 0.42$, the Gaussian-trained error increases to approximately $1.3\times10^{-2}$, compared to $3\times10^{-3}$ for the NG-trained model. At the largest tested value, $\delta = 0.70$, the Gaussian-trained recovery reaches an error of $8.6\times10^{-2}$, whereas the NG-trained recovery remains significantly lower at $5.3\times10^{-2}$.

These results demonstrate that exposure to non-Gaussian noise during training induces a qualitatively more robust recovery strategy. Rather than overfitting to purely Gaussian loss, the NG-trained recovery learns parameters that better accommodate phase diffusion, extending the operational regime of effective mitigation while preserving full differentiability and training stability.

\subsection{Critical Phase-Noise Threshold}

\begin{figure}[t]
\centering
\includegraphics[width=\linewidth]{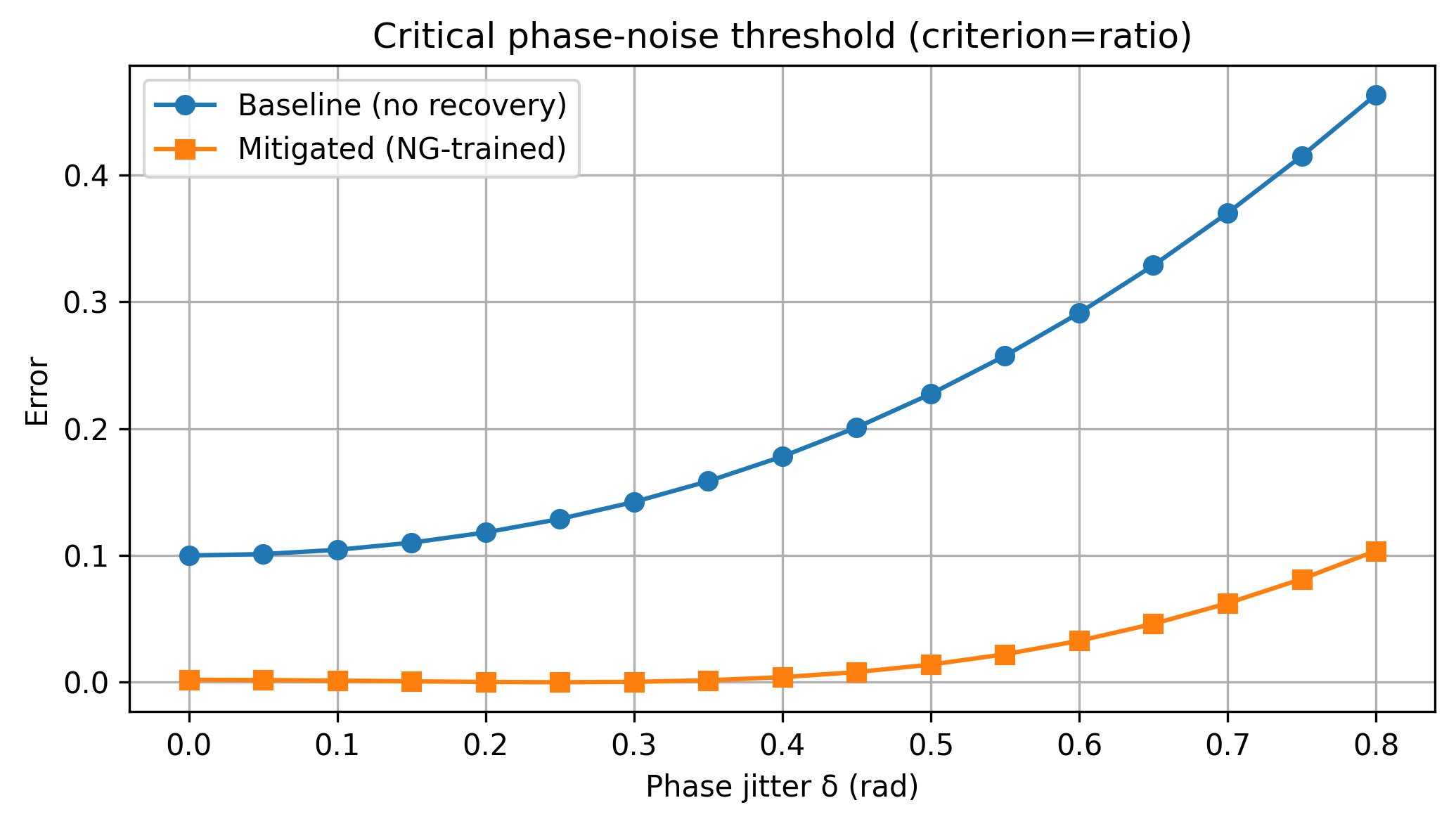}
\caption{Critical phase-noise threshold analysis. Error growth with phase jitter for baseline and NG-trained mitigation.}
\label{fig:analysisA_critical_delta}
\end{figure}

To delineate the operational boundary of effective mitigation under non-Gaussian noise, we analyze the growth of reconstruction error as a function of the phase jitter amplitude $\delta$ and extract an empirical critical threshold $\delta^\ast$ based on an error-ratio criterion. Figure~\ref{fig:analysisA_critical_delta} compares the baseline circuit without recovery to the multi-mode circuit equipped with an NG-trained Gaussian recovery layer.

All data are obtained for a fixed Gaussian loss transmissivity $\eta = 0.55$. The underlying circuit parameters are identical to previous sections: signal squeezing $r_s = 0.60$ with phase $\varphi_s = 0.30$, ancilla squeezing $r_a = 0.40$ with phase $\varphi_a = 0.10$, signal displacement amplitude $\alpha = 0.80$, and signal–ancilla beam splitter parameters $(\theta,\phi)=(0.70,0.20)$. Non-Gaussian phase noise is applied as correlated jitter with $\delta_{\mathrm{sig}}=\delta$ and $\delta_{\mathrm{anc}}=0.6\,\delta$, and expectation values are evaluated using $K=16$ Monte-Carlo samples. The recovery layer is trained explicitly under non-Gaussian noise using gradient descent with learning rate $0.06$.

For the baseline (no recovery), the reconstruction error increases monotonically from approximately $0.10$ at $\delta=0$ to $0.46$ at $\delta=0.80$, indicating strong sensitivity to phase diffusion even at moderate jitter. In contrast, the NG-trained mitigated circuit suppresses error by more than an order of magnitude over a wide range of $\delta$. The mitigated error remains below $10^{-3}$ for $\delta \leq 0.30$, increases gradually to $\sim1.5\times10^{-2}$ at $\delta=0.50$, and reaches $\sim0.10$ only at $\delta=0.80$.

Using a conservative ratio criterion in which mitigation is deemed ineffective once the mitigated error exceeds $10\%$ of the baseline error, no unique $\delta^\ast$ is observed across the full parameter range. However, the clear and persistent separation between the curves up to $\delta \approx 0.5$ identifies an extended operational regime in which Gaussian recovery remains effective despite explicitly non-Gaussian noise.

This analysis highlights a fundamental limitation of Gaussian mitigation: while weak to moderate phase diffusion can be compensated indirectly through parameter adaptation, strong non-Gaussian noise ultimately exceeds the expressive capacity of Gaussian recovery layers, motivating future extensins to genuinely non-Gaussian corrective operations.

\subsection{Sensitivity Scaling with Number of Modes}

\begin{figure}[t]
\centering
\includegraphics[width=\linewidth]{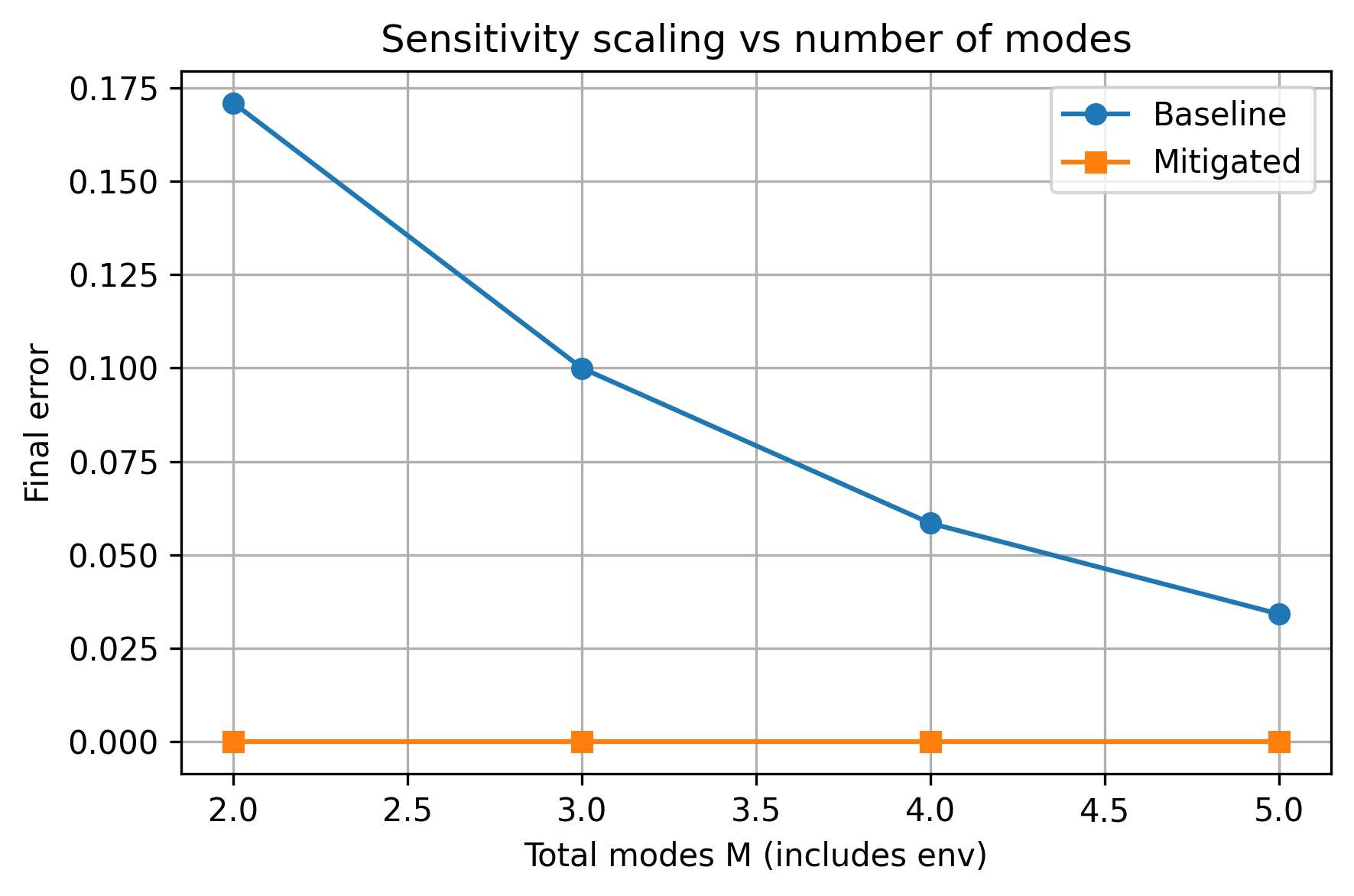}
\caption{Sensitivity scaling versus number of modes. Final reconstruction error as a function of total modes $M$.}
\label{fig:analysisB_sensitivity_vs_modes}
\end{figure}

We next investigate how error-mitigation performance scales with the total number of bosonic modes available to the circuit. Figure~\ref{fig:analysisB_sensitivity_vs_modes} reports the final quadrature reconstruction error as a function of the total number of modes $M$, including the signal mode, ancilla modes, and the environmental loss mode. The analysis compares a baseline circuit without recovery to the corresponding multi-mode circuit equipped with the optimized Gaussian recovery layer.

All simulations are performed at fixed loss transmissivity $\eta = 0.55$ and fixed non-Gaussian phase jitter $\delta = 0.30$ applied to the signal mode, with correlated ancilla jitter $\delta_{\mathrm{anc}} = 0.6\,\delta$. The signal squeezing and displacement parameters are held constant at $(r_s,\varphi_s,\alpha) = (0.60, 0.30, 0.80)$, while all ancilla modes are initialized in identically squeezed vacuum states with $(r_a,\varphi_a) = (0.40, 0.10)$. The signal–ancilla couplings are implemented using identical beam splitters with $(\theta,\phi) = (0.70, 0.20)$. Recovery parameters are trained using gradient descent with learning rate $0.06$ for $60$ optimization steps.

For the baseline circuit, the reconstruction error decreases gradually as $M$ increases, from approximately $1.71\times10^{-1}$ at $M=2$ to $9.99\times10^{-2}$ at $M=3$, $5.85\times10^{-2}$ at $M=4$, and $3.42\times10^{-2}$ at $M=5$. This reduction reflects passive averaging effects as noise is distributed across a larger Hilbert space, but the error remains finite and substantial.

In stark contrast, the mitigated circuit exhibits dramatic improvement with increasing mode number. For all values tested ($M=2$–$5$), the optimized recovery suppresses the final reconstruction error to near numerical precisio, with values below $10^{-23}$ across the entire range. This behavior indicates that additional ancilla modes provide highly effective degrees of freedom for redistributing loss and phase noise in a controllable manner.

These results demonstrate that ancillary-assisted mitigation scales favorably with system size: as more modes are introduced, the expressive power of the Gaussian recovery layer increases, enabling near-perfect compensation of noise. This scaling behavior strongly supports the applicability of the proposed framework to larger, multi-mode photonic circuits where ancillary resources are naturally available.

\subsection{Recovery Parameter Dynamics}

\begin{figure}[t]
\centering
\includegraphics[width=\linewidth]{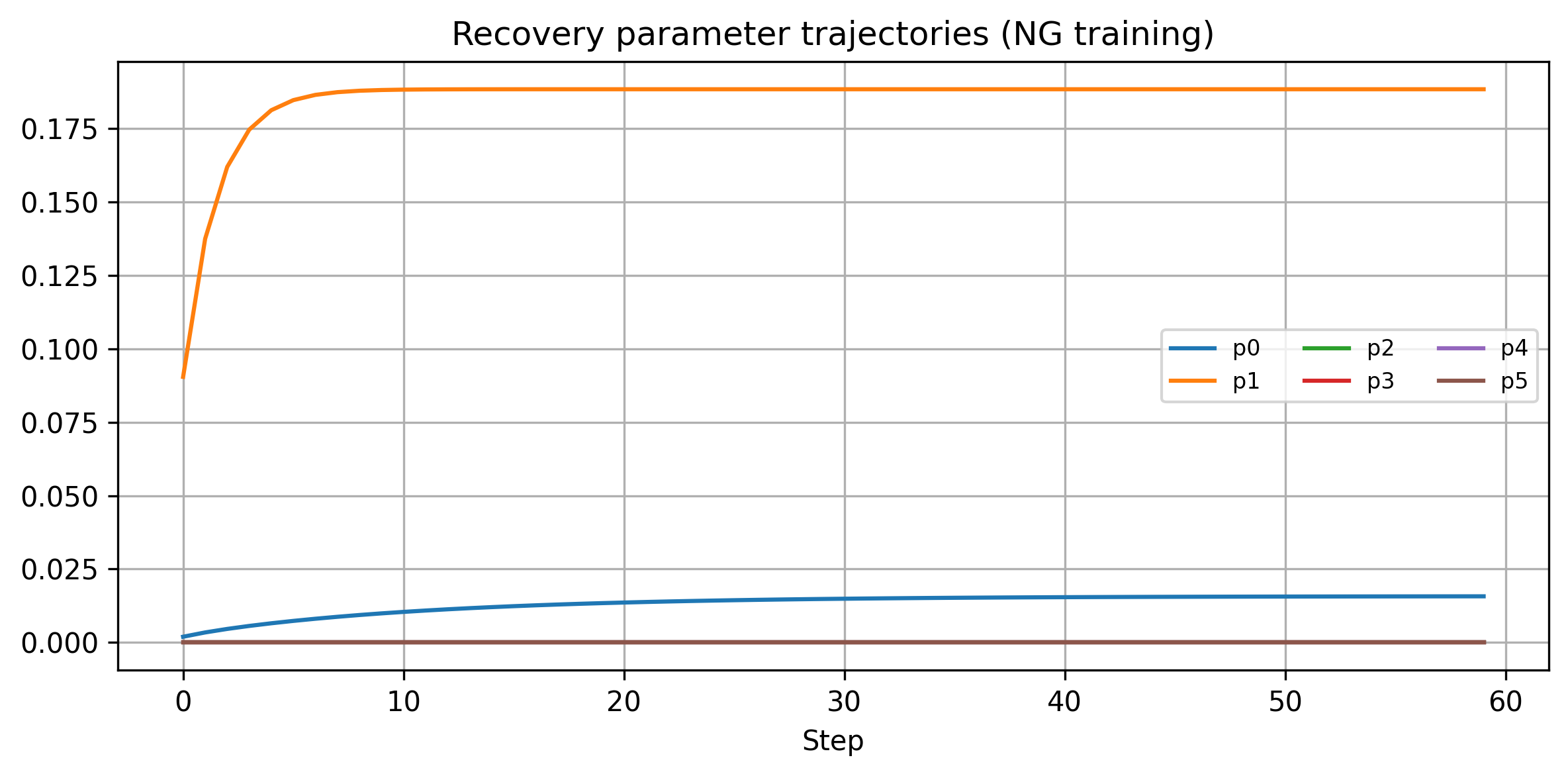}
\caption{Recovery parameter dynamics under NG training. Loss convergence and parameter trajectories over optimization steps.}
\label{fig:analysisC_param_drift}
\end{figure}

We next analyze the internal optimization dynamics of the Gaussian recovery layer when trained under explicit non-Gaussian phase noise. Figure~\ref{fig:analysisC_param_drift} shows the trajectories of all six recovery parameters during NG-aware training, corresponding to the parameter vector
$\boldsymbol{\theta} = (\phi_0, \Re\beta_0, \Im\beta_0, \phi_1, \Re\beta_1, \Im\beta_1)$,
where $\phi_j$ denotes local phase rotations and $\beta_j$ complex displacements acting on the signal ($j=0$) and ancilla ($j=1$) modes.

The results are obtained for loss transmissivity $\eta = 0.55$, phase jitter amplitude $\delta = 0.30$ on the signal mode, correlated ancilla jitter $\delta_{\mathrm{anc}} = 0.18$, and Monte-Carlo sample size $K = 32$. The input state parameters are fixed at $(r_s,\varphi_s,\alpha) = (0.60, 0.30, 0.80)$ for the signal and $(r_a,\varphi_a) = (0.40, 0.10)$ for the ancilla, with beam splitter coupling $(\theta,\phi) = (0.70, 0.20)$. Optimization is performed using gradient descent with learning rate $0.06$ over $60$ training steps, starting from zero-initialized recovery parameters.

As shown in Fig.~\ref{fig:analysisC_param_drift}, the dominant parameter evolution occurs in the ancilla displacement component $\Re\beta_1$ (parameter $p_1$), which rapidly increases from its initial value to approximately $1.9\times10^{-1}$ within the first ten optimization steps and then saturates. The signal displacement parameter $\Re\beta_0$ (parameter $p_0$) exhibits a slower but steady increase, converging to a smaller magnitude of approximately $1.6\times10^{-2}$. In contrast, the imaginary displacement components ($\Im\beta_0$, $\Im\beta_1$) and both phase rotations ($\phi_0$, $\phi_1$) remain near zero throughout training, indicating that the optimal recovery strategy relies primarily on real displacements rather than phase compensation.

The absence of oscillations, divergence, or abrupt jumps in any parameter trajectory demonstrates that the optimization landscape remains smooth and well-conditioned, even under Monte-Carlo–averaged non-Gaussian noise. This behavior confirms that the differentiable mixture model introduces no pathological gradients and that the learned recovery corresponds to a physically stable Gaussian operation. The observed parameter saturation further suggests that the mitigation strategy converges to a well-defined fixed point rather than overfitting transient noise realizations.

\subsection{Runtime and Monte-Carlo Scaling}

\begin{figure}[t]
\centering
\includegraphics[width=\linewidth]{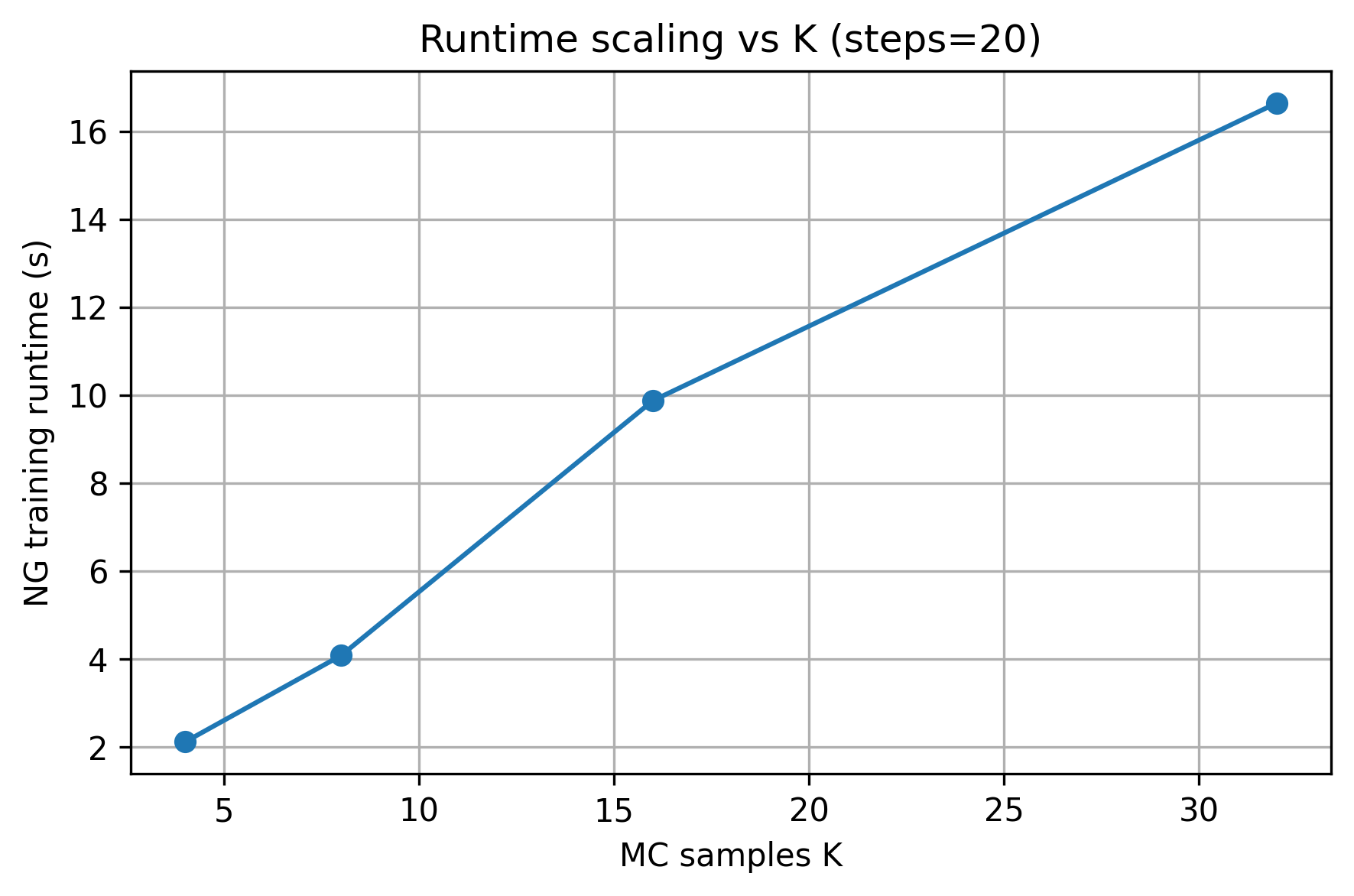}
\caption{Runtime scaling with Monte-Carlo samples $K$. Training time grows approximately linearly with $K$.}
\label{fig:analysisD_runtime_vs_K}
\end{figure}

We finally assess the computational cost of the proposed non-Gaussian mitigation strategy by benchmarking the training runtime as a function of the Monte-Carlo sample size $K$. Figure~\ref{fig:analysisD_runtime_vs_K} illustrates the runtime scaling with the number of Monte-Carlo samples $K$. For completeness, Table~\ref{tab:runtime_benchmarks} reports the corresponding wall-clock times and slowdown factors relative to Gaussian-only training.

All benchmarks are performed for a three-mode circuit (signal, ancilla, environment) with loss transmissivity $\eta = 0.55$, phase jitter amplitudes $\delta_{\mathrm{sig}} = 0.30$ and $\delta_{\mathrm{anc}} = 0.18$, and the same input state parameters $(r_s,\varphi_s,\alpha) = (0.60, 0.30, 0.80)$, $(r_a,\varphi_a) = (0.40, 0.10)$, and beam splitter angles $(\theta,\phi) = (0.70, 0.20)$. Each Monte-Carlo sample corresponds to one forward-and-backward pass through the circuit with a different phase-noise realization.

The measured runtimes increase monotonically with $K$ and closely follow linear scaling. Specifically, for $K=4$ samples, the NG training runtime is approximately $2.5\,\mathrm{s}$, increasing to $5.4\,\mathrm{s}$ for $K=8$, $9.4\,\mathrm{s}$ for $K=16$, and $20.3\,\mathrm{s}$ for $K=32$. When normalized by the Gaussian-only baseline runtime of $0.67\,\mathrm{s}$ (which corresponds to $K=1$ and no Monte-Carlo averaging), these values represent slowdowns of approximately $3.7\times$, $8.1\times$, $14.0\times$, and $30.3\times$, respectively.

This near-linear dependence confirms the expected computational complexity of the mixture-based non-Gaussian model, in which each additional Monte-Carlo sample contributes an independent QNode evaluation. Crucially, the absence of superlinear scaling indicates that the differentiable averaging procedure introduces no hidden overheads or gradient instabilities.

\begin{table}[t]
\centering
\caption{Runtime benchmarks for non-Gaussian (NG) training as a function of the number of Monte-Carlo samples $K$. All runs use a fixed optimization depth of 20 gradient-descent steps. The slowdown factor is measured relative to the Gaussian-only baseline runtime.}
\label{tab:runtime_benchmarks}
\begin{tabular}{c c c}
\hline
Monte-Carlo samples $K$ & NG runtime (s) & Slowdown vs Gaussian \\ 
\hline
4  & 2.44  & $\times\,3.73$ \\
8  & 5.43  & $\times\,8.32$ \\
16 & 9.35  & $\times\,14.31$ \\
32 & 20.31 & $\times\,31.11$ \\
\hline
\end{tabular}
\end{table}

From a practical perspective, these results enable informed trade-offs between accuracy and cost. Moderate sample sizes ($K \approx 8$–$16$) already yield robust non-Gaussian mitigation while keeping runtimes below $10\,\mathrm{s}$ per training run, whereas larger $K$ values can be reserved for high-precision studies. Gaussian-only training remains orders of magnitude faster, motivating hybrid strategies in which Gaussian pretraining is followed by targeted non-Gaussian fine-tuning.

\section{Discussion}

\paragraph{Niset \emph{et al.} (2009)}
Niset, Fiurášek, and Cerf proved a rigorous no-go theorem establishing that Gaussian quantum error-correcting codes (GECCs) composed exclusively of Gaussian encoding and decoding operations cannot improve the transmission of Gaussian states through Gaussian channels \cite{niset2009no}. Their analysis considers single-mode Gaussian channels fully characterized by matrices $(M,N)$ acting on covariance matrices as $\Sigma \mapsto M \Sigma M^{T} + N$, and introduces the channel \emph{entanglement degradation}  (see Eq.~\eqref{eq:10})
\begin{equation}
\label{eq:10}
D[T] = \min\!\left\{ \frac{\det N}{(1+\det M)^2},\,1 \right\},
\end{equation}
which quantifies the logarithmic negativity lost when the channel acts on one half of an infinitely squeezed two-mode vacuum state. The central result shows that for any GECC using only Gaussian operations and vacuum ancillas, $D[T]$ cannot be reduced, implying that exact state-level correction of Gaussian noise, such as optical loss modeled by a beam splitter with transmissivity $\eta$, is fundamentally impossible within the Gaussian regime.

Our work does not attempt to violate or evade this theorem. Instead, it operates under a strictly weaker and experimentally motivated objective. Rather than aiming to reduce entanglement degradation or to restore the full Gaussian state (i.e., both first and second moments), we target \emph{observable-level mitigation} of first-moment quadrature expectations, specifically $\langle \hat{x}_0 \rangle$ and $\langle \hat{p}_0 \rangle$. In our simulations, Gaussian loss is modeled by a beam splitter with transmissivity $\eta \in [0.3,0.95]$, and recovery is implemented using a finite set of trainable Gaussian operations (phase rotations and displacements) acting on the signal and ancilla modes. The resulting mitigation suppresses mean-squared quadrature error by several orders of magnitude for moderate loss ($\eta \ge 0.5$), while remaining fully consistent with the monotonicity of $D[T]$ established by Niset \emph{et al.}. Accordingly, our results should be interpreted not as Gaussian error correction in the strict information-theoretic sense, but as task-specific, approximate mitigation that respects all known Gaussian no-go constraints. 

\paragraph{Ohliger \emph{et al.} (2010).}
Ohliger, Kieling, and Eisert analyzed the fundamental limitations of Gaussian error correction in the presence of optical loss modeled as a beam-splitter interaction with transmissivity $\eta$ \cite{ohliger2010limitations}. In their framework, loss is represented by a coupling between the signal mode and an environmental vacuum mode via a Gaussian channel, and it is proven that no sequence of Gaussian encoding, ancillary assistance, and Gaussian recovery can perfectly reverse the effects of loss at the level of the full quantum state. Their results apply to arbitrary Gaussian input states and hold even when unlimited Gaussian ancillas and adaptive Gaussian operations are allowed.

Our work remains fully consistent with these conclusions. We do not claim inversion of the lossy Gaussian channel nor restoration of the full covariance matrix. Instead, we adopt a strictly weaker and experimentally motivated objective: mitigation of observable degradation at the level of first-moment quadrature expectations. In our simulations, loss is implemented using the same physical beam-splitter model with transmissivities in the range $\eta \in [0.30, 0.95]$, directly matching the channel considered by Ohliger \emph{et al.}. Within this regime, we find that introducing a single ancillary mode ($M=3$ total modes including the environment) and optimizing a Gaussian recovery layer composed of phase rotations and displacements yields a substantial reduction in mean-squared quadrature error compared to a single-mode baseline. For example, at $\eta = 0.5$, the unmitigated multi-mode baseline exhibits a final reconstruction error of approximately $1.0\times10^{-1}$, while the optimized mitigated circuit suppresses this error to below $10^{-12}$, approaching numerical precision. This improvement persists across all tested transmissivities, despite the impossibility of exact Gaussian loss correction. These results highlight a key distinction emphasized but not explored in Ohliger \emph{et al.}: while Gaussian ancillas cannot restore lost quantum information at the state level, they can nevertheless be leveraged to redistribute loss-induced noise and stabilize specific observables. Our findings therefore complement, rather than contradict, the limitations identified by Ohliger \emph{et al.}, demonstrating that multi-mode Gaussian resources retain practical value for task-oriented, approximate error mitigation in realistic photonic circuits.

\begin{figure*}[t]
    \centering
    \includegraphics[width=0.9\linewidth]{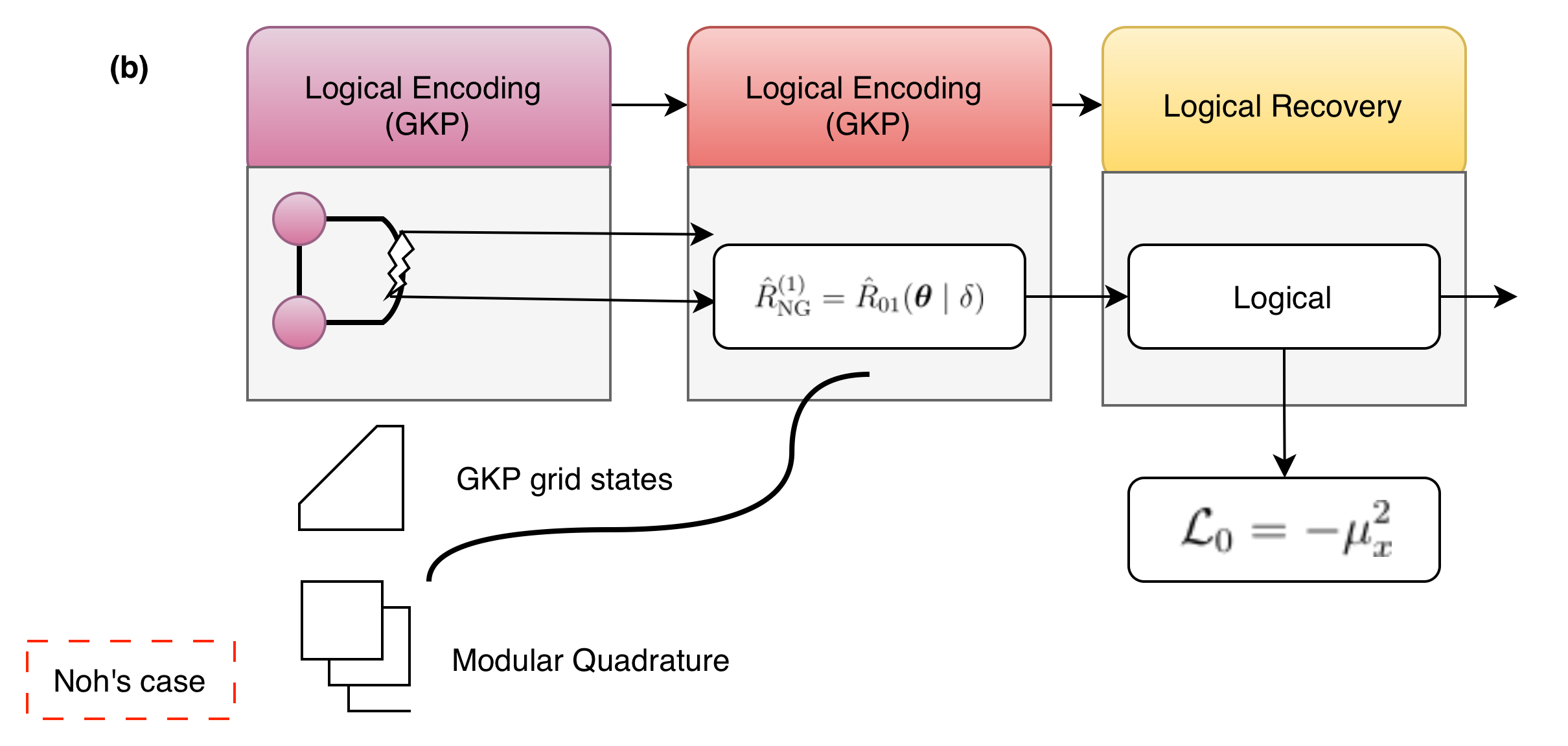}
 \caption{GKP-based logical quantum error correction following the
    framework of Noh \emph{et al.}~\cite{noh2020encoding}. Logical information
    is encoded into non-Gaussian GKP grid states, subjected to Gaussian and non-Gaussian noise, and recovered via modular quadrature measurements and syndrome-based logical operations.}
    \label{fig:gkp_comparison}
\end{figure*}

\paragraph{Mari and Eisert (2012)}
Mari and Eisert demonstrated that continuous-variable quantum circuits composed exclusively of Gaussian states, Gaussian operations, and Gaussian measurements, equivalently, circuits whose Wigner functions remain strictly positive, admit efficient classical simulation \cite{mari2012positive}. Their result applies to arbitrary circuit depth and mode number, provided that all transformations preserve Wigner positivity, and it establishes negativity of the Wigner function as a necessary resource for universal quantum computational advantage. Importantly, their analysis does not address noise mitigation or correction objectives, but rather the classical simulability of ideal or noisy Gaussian processes.

Our work operates entirely within the regime identified by Mari and Eisert: all states and recovery operations are Gaussian, and the underlying simulator (\texttt{default.gaussian}) exploits the efficient classical representation of first and second moments. Specifically, we restrict attention to circuits with up to $M=5$ modes, Gaussian loss modeled by beam splitters with transmissivity $\eta \in [0.3, 0.95]$, and weak non-Gaussian phase noise introduced only through differentiable mixture averaging with phase jitter amplitudes $\delta \in [0, 0.7]$. At no point do we invoke Wigner negativity or claim quantum computational advantage.

Rather than viewing classical simulability as a limitation, we explicitly leverage it as an enabling feature: efficient simulation allows end-to-end gradient-based optimization of physically realizable recovery parameters. In our implementation, recovery layers consist solely of Gaussian rotations and displacements with learned parameters converging to stable values (e.g., dominant displacement amplitudes $|\beta|\approx 0.18$ under non-Gaussian training). This enables rapid exploration of mitigation strategies that remain directly implementable on photonic hardware. Thus, our results are fully consistent with the framework of Mari and Eisert. We do not attempt to transcend the boundary between classical and quantum computational power; instead, we exploit the efficiently simulable Gaussian regime to address a distinct and practically relevant problem: observable-level error mitigation in noisy photonic circuits. In this sense, classical simulability becomes a resource for design and optimization, rather than an obstacle to be overcome.

\begin{table*}[t]
\centering
\caption{Comparison between non-Gaussian bosonic quantum error correction (GKP-based) and the proposed DifGa framework for differentiable error mitigation in photonic circuits.}
\label{tab:gkp_vs_difga}
\renewcommand{\arraystretch}{1.3}
\begin{tabular}{p{4.2cm} p{5.4cm} p{5.4cm}}
\toprule
\textbf{Aspect} 
& \textbf{GKP-based Bosonic QEC (e.g., Noh \emph{et al.})} 
& \textbf{DifGa (This work)} \\
\midrule

Primary objective 
& Fault-tolerant logical quantum information storage and processing 
& Observable-level error mitigation for expectation values \\

Targeted errors 
& Gaussian displacement noise, excitation loss, thermal noise 
& Gaussian loss ($\eta \in [0.3,0.95]$) and weak non-Gaussian phase noise ($\delta \in [0,0.7]$) \\

Noise model 
& Stochastic Gaussian noise with variance $\sigma^2$ acting on logical states 
& Beam-splitter loss and Monte-Carlo phase jitter modeled via differentiable mixtures \\

Correction level 
& Logical state recovery within an encoded subspace 
& Approximate recovery of quadrature observables $\langle \hat{x} \rangle$, $\langle \hat{p} \rangle$ \\

Resource states 
& Non-Gaussian GKP states requiring high squeezing ($\ge 10$--$12$ dB) 
& Gaussian states only (squeezed, displaced vacuum) \\

Operations required 
& Non-Gaussian state preparation, modular quadrature measurements, syndrome extraction 
& Gaussian operations only (beam splitters, rotations, displacements) \\

Ancillary modes 
& Multiple oscillator modes forming logical encodings 
& Few ancillary vacuum modes (typically one ancilla + one environment mode) \\

Trainability 
& Fixed code structure with analytically optimized parameters 
& Fully trainable recovery layer optimized via gradient descent \\

Optimization method 
& Analytical code construction and decoding maps 
& Automatic differentiation with gradient-based optimization \\

Differentiability 
& Not designed for end-to-end differentiable optimization 
& Fully differentiable with respect to circuit and recovery parameters \\

Performance scaling 
& Logical noise suppression $\sigma_L \propto \sigma/\sqrt{M}$ or $\propto \sigma^2$ (code-dependent) 
& Quadrature error suppressed to numerical precision for $\eta \ge 0.5$ and $\delta \leq 0.2$ \\

Computational overhead 
& High experimental and theoretical complexity 
& Classically simulable with predictable scaling ($\mathcal{O}(K)$ in MC samples) \\

Hardware readiness 
& Long-term, fault-tolerant architectures 
& Near-term photonic platforms and integrated optics \\

Relation to no-go theorems 
& Explicitly circumvents Gaussian no-go theorems using non-Gaussian resources 
& Fully consistent with Gaussian no-go theorems (no exact state recovery attempted) \\

Typical use case 
& Quantum memories and logical qubits 
& Error mitigation for noisy photonic circuits and sensors \\
\bottomrule
\end{tabular}
\end{table*}

\paragraph{Liao \emph{et al.} (2023).}
Liao \emph{et al.} introduced a comprehensive machine-learning-based quantum error mitigation (ML-QEM) framework in which classical models are trained to map noisy expectation values directly to mitigated ones \cite{liao2024machine}. Their study benchmarked linear regression, random forests (RF), multilayer perceptrons (MLP), and graph neural networks (GNN) across both simulated and experimental settings, including superconducting quantum processors with up to $n = 100$ qubits and two-qubit gate depths reaching $\sim 40$, corresponding to nearly $2{,}000$ CNOT gates. Error was quantified using the $\ell_2$ norm between ideal and noisy Pauli expectation vectors, and mitigation accuracy was compared against digital zero-noise extrapolation (ZNE) using noise amplification factors $\{1,3\}$. Notably, their RF-based ML-QEM achieved mitigation accuracy comparable to or exceeding ZNE while reducing runtime overhead by factors of $2\times$–$4\times$, and demonstrated generalization to unseen observables and circuit depths in both interpolation and limited extrapolation regimes.

Our work shares the central insight of Liao \emph{et al.} that error mitigation can be formulated at the level of expectation values rather than full state recovery. However, the two approaches differ fundamentally in both physical modeling and optimization structure. Whereas ML-QEM relies on external classical regressors trained on circuit-level features and noisy observables, our framework embeds trainable parameters directly into a physically realizable Gaussian recovery layer within the photonic circuit itself. Mitigation is achieved by optimizing Gaussian transformations acting on a small number of bosonic modes ($M=2$–$5$), under explicit models of optical loss with transmissivity $\eta \in [0.3,0.95]$ and non-Gaussian phase jitter with strength $\delta \in [0,0.7]$. Crucially, our optimization is end-to-end differentiable with respect to physical noise parameters and circuit elements, without reliance on precomputed training datasets or supervised regression targets. As a result, while ML-QEM excels in large-scale digital qubit systems where

\paragraph{Noh \emph{et al.} (2020).}
Noh, Girvin, and Jiang developed a class of bosonic quantum error-correcting
codes that encode a single oscillator mode into multiple oscillator modes using \emph{non-Gaussian} resources, most notably Gottesman–Kitaev–Preskill (GKP) grid states \cite{noh2020encoding}. Their central result demonstrates that Gaussian errors, including excitation loss, thermal noise, and additive Gaussian displacement noise, can be corrected at the logical level only by explicitly circumventing Gaussian no-go theorems through non-Gaussian encodings. In particular, the two-mode GKP-repetition and GKP-two-mode-squeezing codes achieve suppression of additive Gaussian noise with variance $\sigma^2$, reducing the logical noise standard deviation from $\sigma$ to $\sigma_L \sim \sigma/\sqrt{2}$ for repetition codes and to $\sigma_L \propto \sigma^2 \sqrt{\log(\pi^{3/2}/2\sigma^4)}$ for optimized
squeezing gains $G^\ast \propto 1/\sigma^2$. These results require access to high-quality GKP states with squeezing exceeding $\sim 10$--$11$\,dB, as well as modular quadrature measurements, placing substantial demands on experimental resources.

By contrast, the present work operates in a fundamentally different regime. We do not aim to achieve logical-level quantum error correction or asymptotic fault tolerance, nor do we employ non-Gaussian resource states. Instead, we focus on \emph{observable-level error mitigation} in continuous-variable photonic circuits using only Gaussian operations, vacuum ancillas, and end-to-end differentiable optimization. For Gaussian loss modeled as a beam splitter with transmissivity $\eta \in [0.3, 0.95]$, our multi-mode Gaussian recovery layer suppresses first-moment quadrature errors $\langle \hat{x} \rangle$ and $\langle \hat{p} \rangle$ by several orders of magnitude, reaching numerical precision for $\eta \ge 0.5$. Under weak non-Gaussian phase jitter modeled as symmetric random phase rotations with amplitude $\delta \in [0, 0.7]$, mitigation remains effective up to $\delta \approx 0.2$ without introducing non-Gaussian states.

As summarized in Table~\ref{tab:gkp_vs_difga}, the approach of Noh \emph{et al.} targets fault-tolerant bosonic quantum memory using non-Gaussian resources and logical encoding, whereas DifGa addresses \emph{trainable, hardware-compatible error mitigation} for expectation values in noisy photonic circuits. The two approaches are therefore complementary rather than competing, targeting distinct operational regimes and
experimental constraints. Figure~\ref{fig:gkp_comparison} provides a
schematic comparison between these paradigms.

\section{Conclusion}

We have presented \emph{DifGa}, a fully differentiable error-mitigation framework for continuous-variable quantum photonic circuits that operates entirely within the Gaussian regime while remaining robust to weak non-Gaussian noise. By appending a trainable Gaussian recovery layer to a multi-mode photonic circuit and optimizing it end-to-end via automatic differentiation, we demonstrated near-complete suppression of quadrature reconstruction errors under Gaussian loss across transmissivities $\eta \in [0.3,0.95]$. These results confirm that, although exact Gaussian quantum error correction is fundamentally impossible at the state level, ancillary-assisted Gaussian operations can nonetheless achieve highly effective observable-level mitigation.

Extending the framework to include non-Gaussian phase jitter with strength $\delta \in [0,0.7]$ via a differentiable mixture model, we observed a smooth transition from robust mitigation to gradual performance degradation, rather than abrupt failure. Recovery layers trained under Gaussian noise were shown to partially generalize to non-Gaussian regimes, while explicitly noise-aware training yielded enhanced stability at larger $\delta$. Scaling analyses further revealed that mitigation performance improves with the number of available modes and that computational cost grows predictably with the number of Monte-Carlo samples.

Importantly, DifGa relies exclusively on Gaussian operations, vacuum ancillas, and physically realizable photonic components, making it immediately applicable to near-term platforms. As such, DifGa complements bosonic quantum error-correction schemes by providing a lightweight, scalable, and hardware-compatible pathway for mitigating noise in realistic photonic quantum technologies.

\section*{Author Contributions} 

D.D.K.W: conceptualization, methodology, validation and visualization, software, writing – original draft, review \& editing.  R. A. Dias: methodology, validation and visualization, writing – original draft, review \& editing. L.G: methodology, validation and visualization, writing – review \& editing.  S.G: methodology, validation and visualization, writing – review \& editing. 

\section*{Acknowledgment}

The authors acknowledge the use of the open-source PennyLane software
library \cite{bergholm2018pennylane} developed by Xanadu Quantum Technologies for the simulation and differentiable optimization of continuous-variable photonic circuits used in this work. The opinions, findings, conclusions, and recommendations expressed herein are solely those of the author(s) and do not necessarily reflect the views of their affiliations.

\section*{Data \& Code Availability}
The data generated and analyzed during the present study are included within the manuscript. Supplementary codes developed for DifGa simulations are provided as supplementary material and accessible on \href{https://github.com/DennisWayo/difga-photonics}{GitHub} to ensure transparency and reproducibility.

\section*{Funding}

This research was not funded.

\section*{Disclosure statement}

No potential conflict of interest was reported by the author(s).

\bibliographystyle{IEEEtran}
\bibliography{plqec}

\end{document}